\documentclass[twocolumn]{openjournal}
\usepackage{xcolor}
\usepackage{textgreek}
\usepackage[utf8]{inputenc}
\usepackage[english]{babel}
\usepackage{natbib}

\usepackage{hyperref, rotating, booktabs}
\hypersetup{
    unicode, 
    colorlinks=true,
    linkcolor=linkcolor,
    citecolor=linkcolor,
    filecolor=linkcolor,
    urlcolor=linkcolor,
}
\usepackage{cleveref}
\usepackage{color,colortbl}
\definecolor{linkcolor}{rgb}{0.0,0.3,0.5}
\usepackage{tensind}
\tensordelimiter{?}
\DeclareGraphicsExtensions{.bmp,.png,.jpg,.pdf}
\usepackage{verbatim}
\usepackage[normalem]{ulem}
\usepackage{orcidlink}
\usepackage{soul}

\graphicspath{ {./figs/} }

\begin{document}
\title{The Long-Period Radio Transient and Cataclysmic Variable ASKAP 
J1745-5051: Evidence for a $15{,}000$~K White Dwarf and a Sub-Stellar Donor}

\author{Christian Knigge\orcidlink{0000-0002-1116-2553}}
\email{c.knigge@soton.ac.uk}
\affiliation{School of Physics and Astronomy, University of Southampton, Highfield, Southampton SO17 1BJ, UK}

\author{Simone Scaringi\orcidlink{0000-0001-5387-7189}\altaffilmark{$^{\dagger}$}}
\email{simone.scaringi@durham.ac.uk}
\affiliation{Centre for Extragalactic Astronomy, Department of Physics, Durham University, South Road, Durham, DH1 3LE}
\affiliation{INAF-Capodimonte Astronomical Observatory, Salita Moiariello 16, 80131 Naples, IT}

\author{Noel Castro Segura\orcidlink{0000-0002-5870-0443}\altaffilmark{$^{\dagger}$}}
\email{noel.castro-segura@warwick.ac.uk}
\affiliation{Department of Physics, University of Warwick, Gibbet Hill Road, Coventry CV4 7AL, UK}

\author{Domitilla de Martino\orcidlink{0000-0002-5069-4202}}
\email{domitilla.demartino@inaf.it}
\affiliation{INAF-Capodimonte Astronomical Observatory, Salita Moiariello 16, 80131 Naples, IT}

\author{Martina Veresvarska\orcidlink{0000-0002-0146-3096}}
\email{mveresvarska@ice.csic.es}
\affiliation{Institute of Space Sciences (ICE, CSIC), Campus UAB, Carrer de Can Magrans s/n, E-08193 Barcelona, Spain}
\affiliation{Institut d'Estudis Espacials de Catalunya (IEEC), 08860 Castelldefels (Barcelona), Spain}

\altaffiltext{$\dagger$}{These authors contributed comparably to this project.}

% \author{Author\orcidlink{0000-0000-0000-0000}}
% \email{author@you.com}
% \affiliation{Your nice institute / university}
% \affiliation{Your second nice institute / university}

\begin{abstract}
Long-period transients (LPTs) are radio sources that exhibit highly polarized periodic radio bursts on time-scales of minutes to hours. At least some LPTs are associated with white dwarfs (WDs) in close binary systems. However, the evolutionary connection -- if any -- between LPTs and accreting WDs (aka ``cataclysmic variables'' [CVs]) has been unclear. Against this background, the recent discovery of ASKAP J174508.9-505149 has been a breakthrough: this system is a bona-fide LPT that is also an X-ray emitting magnetic CV (mCV) with an orbital period of $P_{orb} \simeq 1.37$ hrs. Here, we construct the broad-band far-ultraviolet through near-infrared spectral-energy distribution (SED) for the system and show that this is well described by two components: a $T_{eff} \simeq 15{,}000$~K WD (which dominates the far-UV through optical bands) and a sub-stellar ($M_{2} \simeq 0.05~M_{\odot}$, $T_{eff} \simeq 1800$~K) donor star (which dominates in the K$_s$ band). Our SED-fitting results differ significantly from those in the discovery paper for four reasons: (i) we fix an issue with the treatment of reddening/extinction; (ii) we discard photometric measurements that are irreparably contaminated by an unrelated star located just 0.9\arcsec\ from  ASKAP J174508.9-505149; (iii) we add near-infrared brightness measurements obtained from PSF-fitting photometry on archival VISTA/VHS observations; (iv) we fit the data with synthetic spectra based on model atmospheres (rather than with blackbodies). The inferred WD temperature is reasonable for an accretion-heated primary in a short-period mCV. The sub-stellar nature of the donor suggests that the system has already evolved past the CV period minimum, i.e. that it is a ``period bouncer''. The SED fit also yields a distance of $d \simeq 320$~pc, only $\simeq 4\times$ larger than that to the nearest confirmed mCV. Since the beaming fraction (the fraction of the sky swept out by the radio beam) is likely to be small, systems like ASKAP J174508.9-505149 could actually make up a large percentage of mCVs. This may point towards a connection between LPTs and the long-standing problem of the ``missing'' population of period bouncers among CVs.
\end{abstract}

% Write your keywords here

\maketitle

\section{Introduction}

Long-Period Transients (LPTs) are radio sources that exhibit highly polarized periodic radio bursts on time-scales of minutes to hours. Since the discovery of the proto-type GLEAM-X J1627-5235 \citep{Hurley-Walker_LPT:2024ApJ...976L..21H}, about a dozen LPTs have been identified \citep{Rea_LPT_review:2026JHEAp..5200566R}. However, their physical nature remains unclear. Even though polarized radio pulses are commonly seen in pulsars and magnetars, the typical recurrence time-scales in those systems are sub-second. And while the time-scales seen in LPTs are similar to those found in ``WD pulsars'' (see below), LPTs can be significantly more luminous and polarized than those systems. 
Despite these challenges, scenarios involving slow-spinning magnetars and highly magnetic WDs are generally favoured. However, much more exotic scenarios have been proposed as well, including models involving primordial black holes, self-lensed pulsar/black-hole binaries, intermediate mass black holes and strange stars \citep[for a review, see][]{Rea_LPT_review:2026JHEAp..5200566R}. 

Models involving magnetic WDs are particularly appealing, since at least some LPTs are associated with WD binary systems. More specifically, before the discovery of ASKAP J174508.9-505149 (referred to as ASKAP J1745-5051 hereafter), only two LPTs had been firmly established as such on the basis of their spectral energy distribution and radial velocity (RV) variations: GLEAM-X J0704-36 \citep{Hurley-Walker_LPT:2024ApJ...976L..21H,Rodriguez_LPT:2025A&A...695L...8R} and ILT J1101+5521 \citep{deRuiter_LPT:2025NatAs.tmp...78D}. Two additional LPTs are strong WD binary candidates without RV confirmation: ASKAP J1448-6856 \citep{Anumarlapudi:2025MNRAS.542.1208A} and CHIME J1634+44 \citep{Bloot:2025A&A...699A.341B}. The optical/infrared counterparts of the remaining LPTs are either unknown or uncertain \citep[e.g.][]{Pelisoli_LPT_UL:2025MNRAS.544L..76P,Rea_LPT_review:2026JHEAp..5200566R}. 

Prior to the recognition that (some) LPTs are associated with WD binaries, two flavours of radio-emitting magnetic WDs in close binary systems were known: one being accreting (the ``magnetic cataclysmic variables'', mCVs), the other non-accreting \citep[the ``WD pulsars'';][]{Marsh2016Natur.537..374M,Pelisoli2023NatAs...7..931P,CastroSegura_WDP3:2025MNRAS.543.2116C}. In mCVs, the WD accretes material from a Roche-lobe-filling companion. This material is channelled onto the WD surface along magnetic field lines, either directly from the inner Lagrangian point (the so-called ``polars'') or from the inner edge of a truncated accretion disk \citep[the ``intermediate polars''; for a review of CVs, see e.g.][]{Warner1995cvs..book.....W}. In polars, the WD spins synchronously with the orbit, i.e. $P_{spin} = P_{orb}$ (with a few systems known as asynchronous polars out of synchronism by a few percent); in intermediate polars, $P_{spin} < P_{orb}$. Radio emission from mCVs is reviewed by \cite{Barrett_CV_review:2020AdSpR..66.1226B}. In the ``WD pulsars'' -- the proto-type being AR~Sco \citep{Marsh2016Natur.537..374M} -- the donor star does not fill its Roche lobe, so the system is detached \citep{Pelisoli_ARSco_long-term:2022MNRAS.516.5052P}. The radio emission in these systems is thought to be powered by the interaction of the magnetic fields of the WD and its companion \citep{Buckley_ARSco:2017NatAs...1E..29B,Garnavich_ARSco_time_resolved_sp:2019ApJ...872...67G}, however the mechanism behind the pulsing behaviour is still debated \citep[e.g.][]{Geng_ARSco:2016ApJ...831L..10G,Katz_ARSco_reconnection:2017ApJ...835..150K,Takata_ARSco_XMM:2018ApJ...853..106T,Potter_Buckley_ARSco:2018MNRAS.481.2384P,duPlessis2022MNRAS.510.2998D,Barrett_ARSco_SED:2025arXiv250506468B}. An evolutionary scenario proposed by \cite{Schreiber_B_in_WDs:2021NatAs...5..648S} links WD pulsars to magnetic CVs, suggesting that accreting WDs may generate stronger magnetic fields as they age and potentially evolve from intermediate polars through a detached WD pulsar phase into polars. In the original version of this scenario, the emergence of magnetic fields is linked to the onset of crystallization. This, however, is challenged by the high temperatures observed in accreting WDs, since compressional heating may inhibit or delay the emergence of such crystallization-driven magnetic fields \citep{Garnavich_ARSco_UV:2021ApJ...908..195G,Pelisoli_J1912_UV:2024MNRAS.527.3826P,Castro-Tapia_crystallization:2024ApJ...975...63C,Camisassa2024}. An update to the scenario by  \cite{2025A&A...698L..22S} therefore suggests that the appearance of magnetic fields might not be linked to crystallization, but instead directly to the age of the binary. 
Clearly, the evolutionary state of these systems -- and their connection to mCVs -- is still poorly understood. 

% --- add a small paragraph mentioning that Rodriguez thought that LPTs, WD pulsars and by extension CVs are connected evolutionary but the WD T from their LPTs is way too low to be consistent (see 4.2 of WDP3 discovery paper also the follow up from the professor from china) --- 
LPTs have been proposed as possible intermediaries between WD pulsars and polars \citep{Rodriguez_LPT:2025A&A...695L...8R}, but the low temperatures of most of the WDs seen in LPTs -- $T_\mathrm{WD}\simeq5$--$7\times10^3\,\rm K$ -- make this connection non-trivial. Such cool temperatures would require the WD pulsar progenitors to have avoided substantial accretion heating and to have entered the pulsar phase already relatively cold. The warmer WDs observed in the closest known WD pulsars, with $T_\mathrm{WD}\simeq11{,}500\,{\rm K}$, therefore pose a challenge to this picture \citep{Garnavich_ARSco_UV:2021ApJ...908..195G,Pelisoli_J1912_UV:2024MNRAS.527.3826P}. 
In contrast, \citet{Yang:2026ApJ...997..124Y} proposed that some LPTs may instead originate from magnetic WD--MD binaries in the pre-mCV phase. There, asynchronous rotation between the WD and the binary orbit can sustain either unipolar-inductor emission or magnetospheric interaction, with the dominant mechanism determined by the relative magnetic moments of the WD and its companion.

Against this background, the recent discovery of ASKAP J1745-5051 as the first LPT that is also clearly an accreting mCV is extremely significant \citep{Rose2026}. Based on the discovery paper and the follow-up X-ray study by \citet{2026A&A...710L..27I}, the system is most likely a polar mCV (or possibly one of the rare {\em slightly} "asynchronous polars"), since all timing signals -- from radio, optical and X-ray bands -- appear to be consistent with a unique (presumably orbital) period. 
\footnote{We note that a wind accretion scenario can be ruled out. The short orbital period and faintness of the secondary imply that the latter must be a very low mass star or brown dwarf. The wind mass-loss rates of such systems are $\lesssim 10^{-13} \mathrm{M_{\odot}~yr^{-1}}$ \citep{m-dwarf-winds,brown-dwarf-winds}, whereas -- for the distance inferred in Section~\ref{sec:SED_fitting} -- the X-ray flux \citep{2026A&A...710L..27I} alone implies an accretion rate $\gtrsim 3 \times 10^{-12} \mathrm{M_{\odot}~yr^{-1}}$.}
Yet its radio properties also clearly establish it as a bona-fide LPT. ASKAP J1745-5051 therefore promises to be the ``missing link'' between LPTs, mCVs and perhaps WD pulsars. The obvious analogy here is with the ``transitional millisecond pulsars'' (tMSPs), which switch between accretion-powered X-ray-binary (XRB) and rotation-powered radio-pulsar states \citep{Papitto_deMartino:2022ASSL..465..157P}. It was the discovery of the proto-typical tMSP PSR J1023+0038 by \citet{Archibald:2009Sci...324.1411A} that conclusively established the evolutionary connection between XRBs and MSPs. 

\begin{center}
    \begin{figure*}[!t]
        \centering
    	\includegraphics[width=\textwidth]{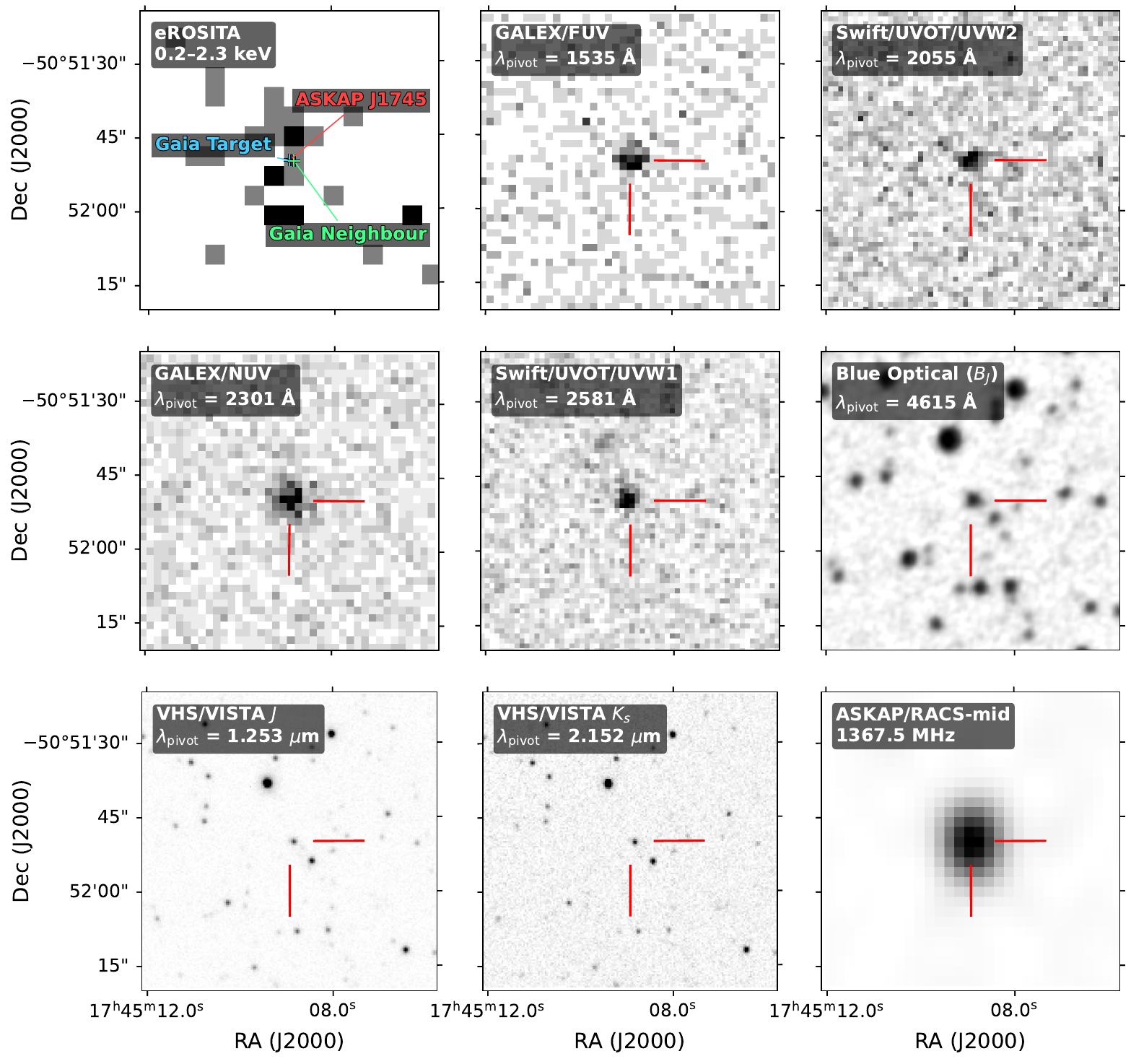}
    	\caption{1\arcmin\ $\times$ 1\arcmin\ finding charts centered in ASKAP J1745-5051 covering wavelengths from X-ray to radio. In the top left panel the position of the ASKAP J1745-5051 and the two {\it Gaia} sources are shown with coloured crosses.}
        \label{fig:finder}
    \end{figure*}
\end{center}

Unfortunately, many of the basic observational properties of ASKAP J1745-5051 remain unknown. For example, while the source has a \textit{Gaia} counterpart, its parallax ($\pi = 1.75 \pm 0.91$~mas) is essentially unconstrained \citep{2023A&A...674A...1G}. Partly as a result of this, the two-blackbody SED fits carried out by \citet{Rose2026} could only place rough constraints on the system component(s) that dominate in the ultraviolet and optical bands. 

\begin{center}
    \begin{figure*}[!t]
        \centering
    	\includegraphics[width=\textwidth]{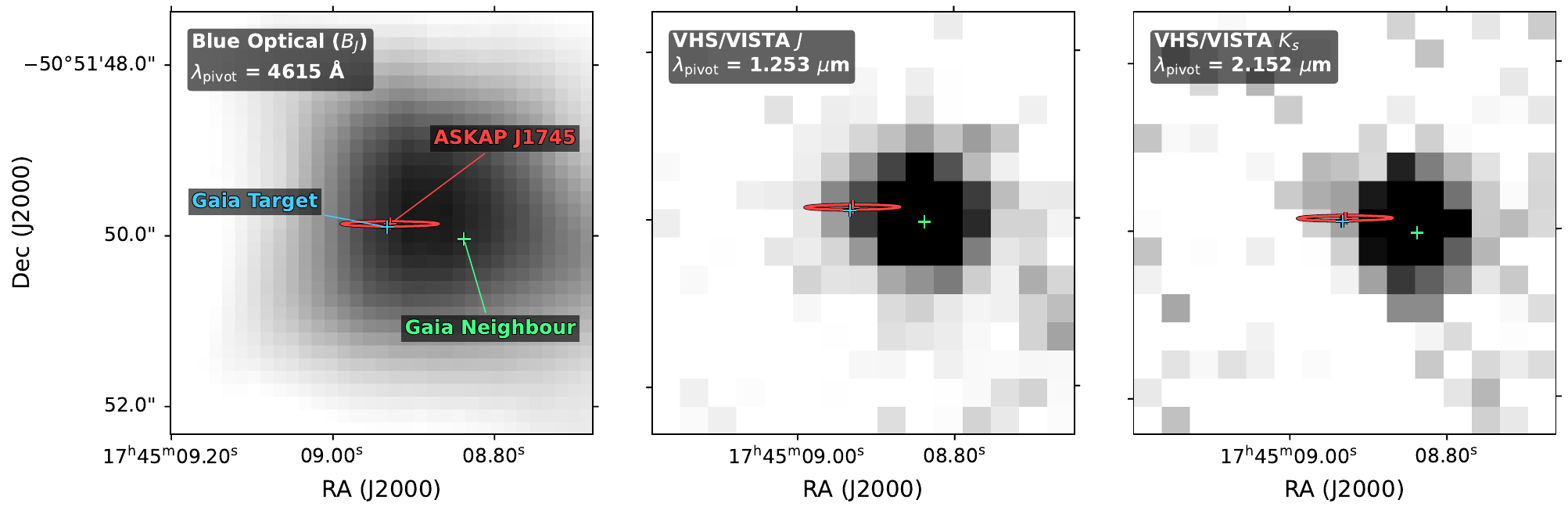}
    	\caption{$5\arcsec \times 5\arcsec$ finding charts from the highest-resolution images centred on ASKAP J1745-5051, showing the positions of the two \textit{Gaia} sources, marked by light green and blue crosses, and the position of the LPT ASKAP J1745-5051 derived from radio observations, indicated by the red ellipse centred on the red cross. The images show clearly how the position of the radio source can be uniquely identified with the \textit{Gaia} counterpart, it also illustrates the blending of the two sources, and demonstrate that the interloper dominates the redder bands.}
        \label{fig:finder_zoom}
    \end{figure*}
\end{center}

The purpose of this paper is to improve on this situation. More specifically, in Section~\ref{sec:SED_construction} we first construct a wider and more reliable broad-band SED for the system. This involves (i) fixing an issue with the reddening corrections applied by \citet{Rose2026} to the ultraviolet and optical fluxes; (ii) discarding photometric measurements based on optical images in which ASKAP J1745-5051 is irreparably blended with an apparently unrelated close neighbour of comparable optical brightness; and (iii) adding near-infrared measurements obtained via PSF-fitting photometry on VISTA/VHS images. In Section~\ref{sec:SED_fitting}, we then fit the resulting SED with a model consisting of two synthetic spectra: a DA WD primary and a low-mass, stellar or sub-stellar secondary. As we shall see, the data is reasonably well described by the combination of a $\simeq 15{,}000$~K WD and a cool ($T \simeq 1700~{\mathrm K} - 2000~{\mathrm K}$) sub-stellar ($M_2 \simeq 0.05~\mathrm{M_{\odot}}$) secondary. In Section~\ref{sec:discussion}, we discuss the implication of these results for understanding ASKAP J1745-5051, specifically, and the connection between LPTs and mCVs, more generally. Finally, in Section~\ref{sec:summary}, we summarize our conclusions.

\section{SED Construction}
\label{sec:SED_construction}

We used the data sources listed in \citet{Rose2026} as a starting point for constructing our broad-band SED of ASKAP J1745-5051. After searching for additional archival observations of the field, we supplemented the \citet{Rose2026} data set with near-IR observations obtained by the VHS Survey \citep{McMahon21}.
Table~\ref{tab:obs_summary} provides a summary of the observations we have used in this work and the corresponding photometric measurements. Finding charts for all wavebands, covering a 1\arcmin\ $\times$ 1\arcmin\ field centered on the target, are provided in Figure~\ref{fig:finder}. 

We note that Table~\ref{tab:obs_summary} lists one set of Swift/UVOT observations (in the near-UV UVW2, UVM2 and UVW1 filters), but there are actually two sets in the archive. As discussed by \citet{Rose2026}, the first (ID00016563001) was much shorter than the second (ID00016563005). Inspection of the images shows that the first observation also suffered from poor tracking.  We therefore only use the second set of observations for the present study.

We do not apply any reddening correction to the observed brightness estimates, choosing instead to deal with extinction as part of our modelling\footnote{One of the first author's mottos is ``Mess with the model, not the data.''}.
The photometric data used by \citet{Rose2026} for their SED fitting is provided in their Supplementary Table~1. This is meant to include a correction for reddening by $E(B-V) = 0.2$. However, their extinction-corrected magnitudes are {\em fainter} than the uncorrected ones. Since they have kindly made their data set and analysis code public, we have been able to trace this to a bug in their extinction-correction routine. The code inadvertently {\em applies} extinction to the data, rather than {\em correcting for} it.

% Observation log table for ASKAP J1745-5051 / LPT-CV
% Requires: booktabs, array
% Usage: \input{obs_table.tex}  (or copy into your document)

\begin{table*}
\centering
\caption{Summary of multi-wavelength observations used in this work.}
\label{tab:obs_summary}
\footnotesize
\setlength{\tabcolsep}{3.5pt}
\renewcommand{\arraystretch}{1.25}
\newcommand{\lp}{\ensuremath{\lambda_{\mathrm{pivot}}}}
\begin{tabular}{
  >{\raggedright\arraybackslash}p{2.6cm}
  >{\centering\arraybackslash}p{1.5cm}
  >{\raggedright\arraybackslash}p{1.3cm}
  >{\raggedright\arraybackslash}p{1.3cm}
  >{\centering\arraybackslash}p{1.7cm}
  >{\centering\arraybackslash}p{1.8cm}
  >{\centering\arraybackslash}p{1.5cm}
  >{\raggedright\arraybackslash}p{3.6cm}}
\toprule
Observatory / Instrument / Survey &
  Band /\newline\hspace*{-5mm} Filter &
  Type &
  Centre &
  Flux / Magnitude &
  Units / System &
  Observation Date&
  Comments \\
\midrule
%
% --- X-ray ---
eROSITA/SRG &
  0.2--2.3\,keV &
  X-ray &
  1 keV&
  $-12.51 \pm 0.08$ &
  \footnotesize{$\log$(erg/s/cm$^{2}$)} &
  2020-03-26 &
  Only used in finder chart. \\[5pt]
%
% --- Ultraviolet ---
GALEX/AIS &
  FUV &
  Far-UV &
  $1535$\,\AA &
  $\phantom{-}19.84 \pm 0.15$ &
  AB &
  2006-08-09 &
  \dots \\[5pt]
Swift/UVOT &
  UVW2 &
  Near-UV &
  $2055$\,\AA &
  $\phantom{-}18.47 \pm 0.16$ &
  Vega &
  2024-05-16 &
  \dots \\[5pt]
Swift/UVOT &
  UVM2 &
  Near-UV &
  $2231$\,\AA &
  $\phantom{-}18.18 \pm 0.15$ &
  Vega &
  2024-05-16 &
  \dots \\[5pt]
GALEX/AIS &
  NUV &
  Near-UV &
  $2301$\,\AA &
  $\phantom{-}19.67 \pm 0.11$ &
  AB &
  2006-08-09 &
  \dots \\[5pt]
Swift/UVOT &
  UVW1 &
  Near-UV &
  $2581$\,\AA &
  $\phantom{-}18.96 \pm 0.20$ &
  Vega &
  2024-05-16 &
  \dots \\[5pt]
%
% --- Optical ---
SERC-J/MAMA &
  $B_J$ &
  Blue \newline optical &
  $4615$\,\AA &
  \dots &
  \dots &
  1975-08-08 &
  UKST SERC-J survey plate J\,1756 (field 228), digitised by the Observatoire de Paris MAMA scanner. Target blended with neighbour. Only used in finder chart \\[5pt]
Gaia/DR3 &
  $G$ &
  Wide optical &
  $6218$\,\AA &
  $\phantom{-}19.45 \pm 0.04$ &
  Gaia &
  \dots &
  Observations span 2014-07-25 to 2017-05-28. Gaia$\to$Vega offset ($\simeq 0.03\,$mag) neglected.\\[5pt]
%
% --- Near-infrared ---
VISTA/VIRCAM/VHS &
  $J$ &
  Near-IR &
  $1.253\,\mu$m &
  $\phantom{-}19.19 \pm 0.10$ &
  Vega &
  2014-07-28 &
  Gaussian-PSF fit\\[2pt]
 &
 &
 &
 &
  $\phantom{-}19.79 \pm 0.16$ &
 &
 &
  Moffat-PSF fit\\[5pt]
VISTA/VIRCAM/VHS&
  $K_s$ &
  Near-IR &
  $2.152\,\mu$m &
  $\phantom{-}18.48 \pm 0.18$ &
  Vega &
  2014-07-28 &
%  \newline Epoch 1 \newline v20140727\_00333  &
  Gaussian-PSF fit\\[2pt]
 &
 &
 &
 &
  $\phantom{-}18.69 \pm 0.21$ &
 &
 T03:08:36&
 Moffat-PSF fit\\[5pt]
VISTA/VIRCAM/VHS&
  $K_s$ &
  Near-IR &
  $2.152\,\mu$m &
  $\phantom{-}18.32 \pm 0.16$ &
  Vega &
  2014-07-28 &
%  \newline Epoch 2 \newline v20140727\_00337 &
  Gaussian-PSF fit\\[2pt]
 &
 &
 &
 &
  $\phantom{-}18.47 \pm 0.18$ &
 &
 T03:10:33&
 Moffat-PSF fit\\[5pt]
%
% --- Radio ---
ASKAP/RACS &
  mid &
  Radio &
  1367.5\,MHz &
  $\phantom{-}22.40 \pm 0.36$ &
  mJy&
  2020-12-27 &
  Only used in finder chart. \\[5pt]
\bottomrule
\end{tabular}
\begin{flushleft}
\smallskip
\end{flushleft}
\end{table*}

\subsection{An Interloper Within 1\texorpdfstring{\arcsec}{"} from  ASKAP J1745-5051}
\label{sec:interlooper}

The \textit{Gaia} DR3 catalogue reveals that ASKAP J1745-5051 (Gaia DR3 ID: 5946454415417964032; $m_G$ = 19.45) is located just $0.9$\arcsec\ from a slightly brighter optical source (Gaia DR3 ID: 5946454411127231488; $m_G$ = 19.33). Figure~\ref{fig:finder_zoom} shows a zoom-in on the central 5\arcsec\ $\times$ 5\arcsec\ region around ASKAP J1745-5051 in our highest resolution (blue optical and near-IR) images. The existence of this neighbour was already noted by \citet{Rose2026}. However, its presence means that the \textit{Gaia} $B_p$ and $R_p$ estimates for ASKAP J1745-5051 are unreliable: the effective spatial resolution of \textit{Gaia} is $\simeq 0.4$\arcsec\ -- 0.7\arcsec\ in the G-band \citep{fabri}, but only 2\arcsec\ - 3\arcsec\ in the $B_p$ and $R_p$ bands \citep{riello}.

The problem can be confirmed quantitatively via \textit{Gaia}'s  \texttt{phot\_bp\_rp\_excess\_factor}
parameter, which is defined as the flux ratio $(F_{B_p}+F_{R_p})/F_{G}$. For a well-behaved isolated source, this ratio should be $\simeq 1$. Values above $\simeq 1.3$ indicate significant blending problems. For the \textit{Gaia} counterpart to ASKAP J1745-5051, \texttt{phot\_bp\_rp\_excess\_factor} = 2.89.
We therefore caution that the location of ASKAP J1745-5051 in the \textit{Gaia} CMD shown by \citet{Rose2026} is unreliable. 
\footnote{\cite{riello} define an improved version of the $B_p-R_p$ excess factor that applies a colour correction, $C^*$, which is designed to fall near $C^* \simeq 0$ for clean, unblended sources. For ASKAP J1745-5051, we obtain $C^* \simeq 1.7$ (using the implementation by Anthony Brown: https://github.com/agabrown/gaiaedr3-flux-excess-correction). This is outside the entire range of Gaia sources shown in Fig. 18 of \citep{riello}, confirming that the $B_p$ and $R_p$ colours are unreliable.} 
We thus do not include the \textit{Gaia} $B_p$ and $R_p$ estimates in our broad-band SED. 
\footnote{The extent to which the published $B_p-R_p$ colour is offset from the true value depends on how similar the intrinsic colour of ASKAP J1745-5051 is to that of its neighbour. The published colours are $B_p-R_p = 1.08$ (ASKAP J1745-5051) and $B_p-R_p = 1.49$ (neighbour). However, since blending will tend to average out the colour differences, these estimates are bound to be  more similar than the true values.}

It is worth noting that there are currently {\em no} optical images of the field in which ASKAP J1745-5051 can be resolved from its neighbour. The MAMA scan of the SERC-J photographic plate shown in Figures~\ref{fig:finder} and \ref{fig:finder_zoom} is likely to be the highest resolution image available, and it is clearly not sufficient to attempt deblending via PSF-fitting. The only modern CCD-based optical images of the field come from the SkyMapper survey, but the PSF in these images has FWHM $\simeq$ 2.5\arcsec\ - 3.2\arcsec \citep{2019PASA...36...33O}. As it stands, the \textit{Gaia} G-band magnitude is the {\em only} reliable optical brightness measurement for ASKAP J1745-5051.

The neighbouring star appears to be unrelated to ASKAP J1745-5051. Neither source has a meaningful parallax, presumably due to the blending problem (ASKAP J1745-5051: $\pi = 1.75 \pm 0.91$~mas; neighbour: $\pi = 1.03 \pm 0.54$~mas). However, they do have very different proper motions (ASKAP J1745-5051: $\mu_\alpha = 2.98 \pm 0.85~\mathrm{mas~yr^{-1}}$, $\mu_\delta = 3.77 \pm 0.75~\mathrm{mas~yr^{-1}}$; neighbour: 
$\mu_\alpha = -1.72 \pm 0.51~\mathrm{mas~yr^{-1}}$, $\mu_\delta = 8.43 \pm 0.44~\mathrm{mas~yr^{-1}}$). These values seem irreconcilable, even if we allow for additional systematic uncertainties associated with blending.

Importantly, despite the presence of the interloper, the association of the radio, X-ray and UV sources with Gaia DR3 5946454415417964032 is secure. As shown in Figure~\ref{fig:finder_zoom}, the position of the radio source is precise enough to uniquely identify the correct \textit{Gaia} counterpart. Since the sky densities of radio, X-ray and UV sources are low, the probability that these sources are chance coincidences is negligible. Moreover, \citet{Rose2026} have detected the period seen in the radio data also in optical spectroscopy, and \citet[][]{2026A&A...710L..27I} have detected the same period also in X-rays.

\begin{center}
    \begin{figure*}[!t]
        \centering
    	\includegraphics[width=\textwidth]{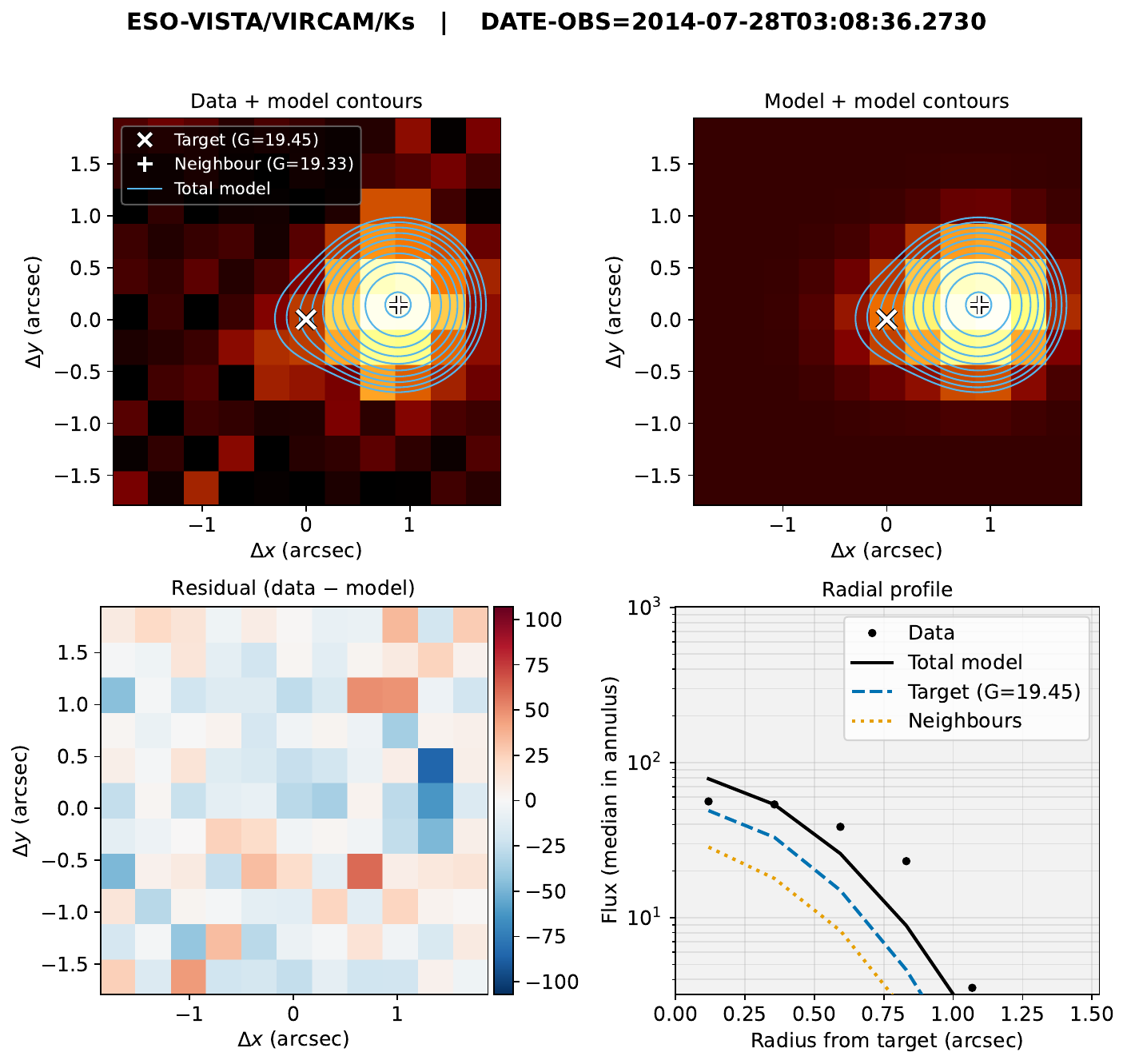}
    	\caption{PSF fitting of the ESO-VISTA/VIRCAM $K_s$ band with two Gaussian components. {\bf Top left:} Data with isocontours of the composite model fit. The fitted position for ASKAP J1745-5051 and the neighbouring star are shown with an $\times$ and $+$ symbol, respectively. {\bf Top right:} model image constructed from the two Gaussians, with the same isocontours overlaid. {\bf Bottom left}: residual image (data-model). {\bf Bottom right:} the median fluxes in each annulus as a function of radius for the data, total model and each individual fit are shown with black dots, white solid line, green dashed line and dotted line, respectively.}
        \label{fig:psf_sanity1}
    \end{figure*}
\end{center}

\begin{center}
    \begin{figure*}[!t]
        \centering
    	\includegraphics[width=\textwidth]{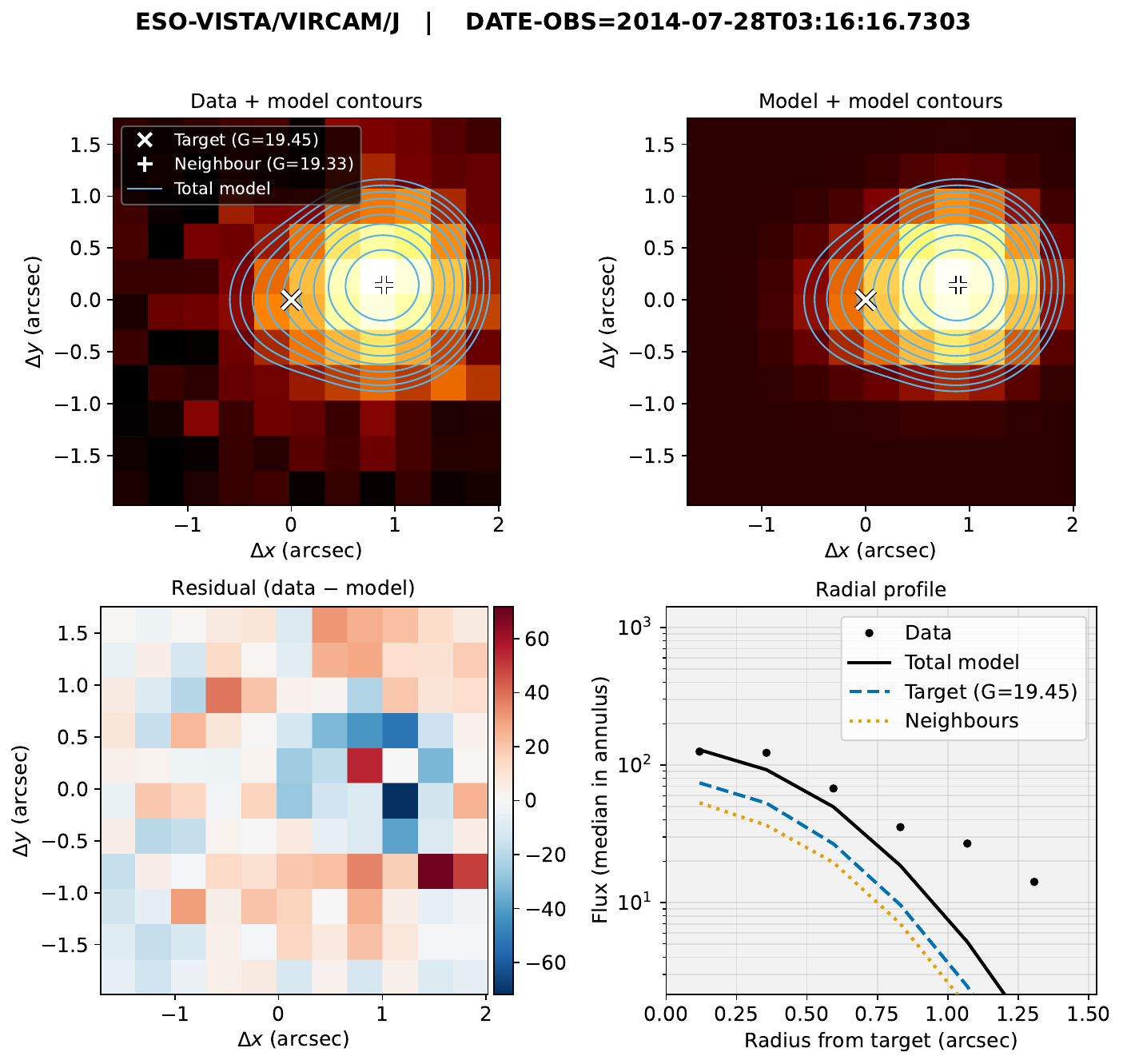}
    	\caption{Same as Figure~\ref{fig:psf_sanity1}, but for ESO-VISTA/VIRCAM $J$ band}
        \label{fig:psf_sanity2}
    \end{figure*}
\end{center}

\subsection{Estimating The Near-IR Brightness of ASKAP J1745-5051}

The VISTA/VHS images have {\em just} enough spatial resolution -- FWHM $\simeq$ 0.9\arcsec\ -- to warrant an attempt at PSF-fitting to estimate the near-IR brightness of ASKAP J1745-5051. We have therefore downloaded the relevant images and carried out PSF-fitting photometry as follows. For each image, we run a source detection algorithm on a 1\arcmin~radius field centered on ASKAP J1745-5051. An isolated sub-set of the detected sources is used to estimate the FWHM of the PSF. The detected sources are cross-matched to the \textit{Gaia} DR3 sources in the same field, and the resulting matches are used to refine the image world-coordinate system (WCS) slightly. We then carry out forced PSF-fitting photometry on the locations of the \textit{Gaia} sources in the field, adopting a Gaussian PSF with the FWHM estimated previously. We finally calibrate the photometry by cross-matching to the VHS catalogue and applying a zero-point that optimally matches our instrumental magnitudes to the VHS magnitudes. Our error budget thus includes contributions from noise in the images, from blending between sources and from the global calibration. 

Figures~\ref{fig:psf_sanity1} and ~\ref{fig:psf_sanity2} provide diagnostic plots that illustrate the quality of these fits for the $J$-band and one $K_s$-band image. It should be clear from these figures that the deblending process is difficult. The neighbour is clearly brighter than ASKAP J1745-5051 in the near-IR, and the 2-D brightness distribution shows only a {\em slight} elongation towards the target position (especially in the $K_s$ band).

As a check on the systematics associated with this PSF-fitting procedure, we repeated the entire process, but this time assuming a Moffat PSF. There are then two PSF shape parameters -- $\alpha$ and $\beta$. As before, these were estimated empirically from isolated sources in the field. Table~\ref{tab:obs_summary} also lists these Moffat-PSF-based near-IR brightness estimates for ASKAP J1745-5051, which differ by 0.2 mag - 0.6 mag from those based on the Gaussian PSF. As expected, the Moffat-PSF-based estimates are systematically fainter, because this PSF has broader wings. As a result, a larger proportion of the flux at the position of ASKAP J1745-5051 is attributed to the neighbour. However, both Gaussian- and Moffat-PSF-fitting approaches produce magnitude estimates with {\em internal} uncertainties of 0.1 mag - 0.2 mag, suggesting that -- regardless of the measurement approach -- ASKAP J1745-5051 is at least marginally detected in these images. We will adopt the magnitude estimates obtained with the Gaussian PSF as our default. However, we will discuss the implications of these systematic uncertainties -- including the possibility that the detections are false positives -- in Section~\ref{sec:discussion}.

\begin{center}
    \begin{figure*}[!t]
        \centering
    	\includegraphics[width=\textwidth]{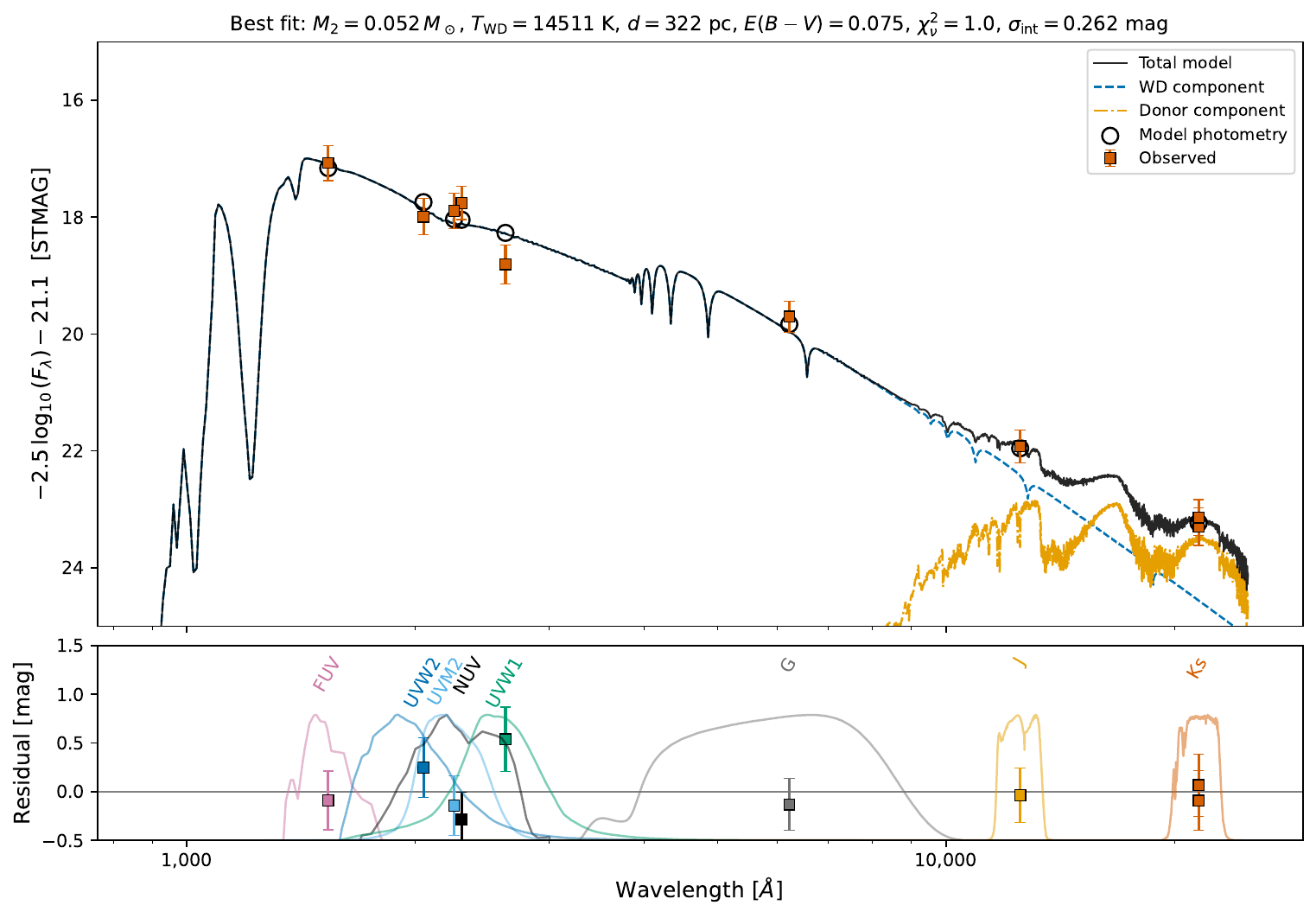}
    	\caption{Spectral energy distribution of ASKAP J1745-5051 constructed using the photometry compiled and derived in this work. The observed fluxes are shown as orange-red squares. The SED is modelled with a DA white dwarf atmosphere and a late-type stellar companion atmosphere, shown by the blue dashed and orange dash-dotted curves, respectively. The resulting composite model is shown by the black solid curve, while the open circles indicate the corresponding synthetic photometry. The transmission function of each band is shown in the bottom panel, together with the fit residuals.}
        
        \label{fig:sed_fit}
    \end{figure*}
\end{center}

\section{SED Fitting}
\label{sec:SED_fitting}

In order to determine the physical properties of the ASKAP J1745-5051 binary system, we have carried out a simple 2-component model fit to the SED. The two components we consider are the WD primary and the low-mass secondary. We do not include any contribution from the accretion flow itself, such as an accretion disk or magnetic accretion curtains. The rationale for and implications of this choice are discussed in Section~\ref{sec:simpleModel}. 

We use \cite{Koester10} synthetic DA WD spectra to model the flux produced by compact primary in ASKAP J1745-5051. We adopt a mass of $M_{WD} \simeq 0.8~M_{\odot}$ for this component, roughly the average mass of WDs in CVs \citep[e.g][]{Knigge2008ApJ...683.1006K,2008MNRAS.388.1582L,2011MNRAS.415.2025S,2019MNRAS.486.5535M,zorotovic2011, 2022MNRAS.510.6110P}. The corresponding  WD radius ($R_{WD} \simeq 7 \times 10^{8}$~cm) and surface gravity ($\log{g} \simeq 8.3$) are calculated from the WD cooling models of \citet{bedard}.
\footnote\url{{https://www.astro.umontreal.ca/~bergeron/CoolingModels/}}
These values are actually slightly temperature-dependent, and we account for this in our fits. 

Since the orbital period of ASKAP J1745-5051 is spectroscopically confirmed -- $P_{orb} = 1.368 \pm 0.053~\mathrm{hr} = 82.1 \pm 3.2~\mathrm{min}$ \citep{Rose2026}  -- the secondary can be placed almost uniquely on the semi-empirical CV donor sequence \citep[][hereafter KBP11]{Knigge2006,Knigge+2011ApJS..194...28K}. The placement is not quite unique, since this $P_{orb}$ is very close to the minimum in the CV orbital period distribution, $P_{min} = 82.4 \pm 0.7$~min \citep{Gaensicke:2009MNRAS.397.2170G, 2019MNRAS.486.5535M}. 

The period minimum exists because the direction of orbital period evolution depends only on the mass-radius index ($\zeta$) of the donor. The critical point, where the direction reverses, corresponds to $\zeta \simeq 1/3$. A CV with a roughly main-sequence donor ($\zeta \simeq 1$) initially evolves from long to short orbital periods. However, the mass-losing donor must eventually reach the Hydrogen-burning limit and become a brown dwarf \citep[e.g. ][]{HernandezSantisteban:2016Natur.533..366H}. Such sub-stellar objects are characterized by $\zeta \simeq 0$. Somewhere along the line therefore -- it turns out very close to the stellar/sub-stellar transition -- the direction of period evolution must turn around. CVs that have already evolved beyond $P_{min}$ are expected to harbour sub-stellar donors and are often referred to as ``post-period-minimum'' systems or, more casually, ``period bouncers''. \footnote{ It is worth noting that, in principle, CVs could also be {\em born} with a brown-dwarf donor at periods below $P_{min}$ \citep{politano-bd, littlefair1507}. Such systems would monotonically evolve towards longer periods, joining the "normal" CV evolution track near $P_{min}$. No such systems have been convincingly identified to date. The best candidate (on the basis that its orbital period is significantly below $P_{min}$) was SDSS J150722.30+523039.8 \cite{littlefair1507}. However, this system is likely to be a low-metallicity halo CV \citep{patterson1507, uthas1507}, which also naturally explains it short orbital period. (Low-metallicity donors are denser, and the period-density relation for donors in Roche-lobe filling systems enforces $P_{orb}\rho_2={\mathrm constant}$ (where $\rho_2$ is the donor density).}

\begin{center}
    \begin{figure*}[!t]
        \centering
    	\includegraphics[width=\textwidth]{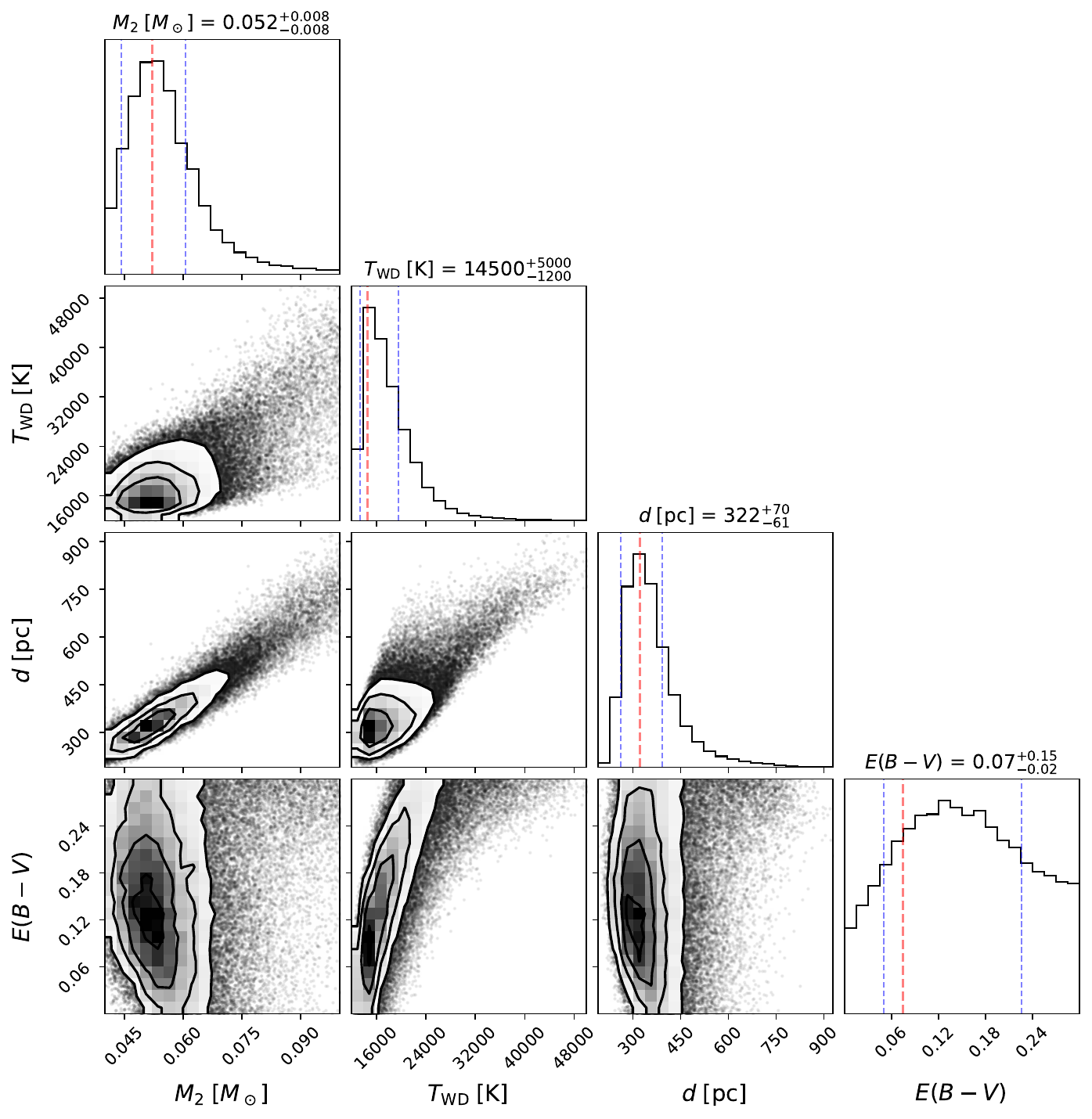}
    	\caption{Posterior distributions of the ASKAP J1745-5051 system parameters inferred from the SED modelling presented in Figure~\ref{fig:sed_fit}. We adopted a WD mass of $M_{WD}= 0.8~M_{\odot}$ with radius $R_{WD}\simeq7\times10^8\mathrm{cm}$ and surface gravity $\log{g}\simeq8.3$. The corner plot shows all fitted parameters except for the intrinsic scatter term, $\sigma_\mathrm{int}$, which was added in quadrature to the uncertainties of all photometric measurements.}
        \label{fig:sed_fit_corner}
    \end{figure*}
\end{center}

ASKAP J1745-5051 orbital period is such that it could either be a pre- or a post-period-minimum system. Taking a $\simeq 2\sigma$ interval around $P_{orb}$, the KBP11 ``optimal'' donor sequence suggests that the allowed range of secondary masses is $0.04 \lesssim M_2/M_{\odot}\lesssim 0.09$. The existence of the donor sequence is a direct consequence of the fact that CVs are semi-detached binaries, and since ASKAP J1745-5051 is a confirmed CV, its donor {\em must} lie close to this sequence. We therefore constrain the component representing the secondary in our SED fit to lie on the KBP11 optimal sequence. For any given $M_2$, this fixes $R_2$ (and hence $\log{g_2}$), but also the donor temperature, $T_2$. 

Since our SED fit extends from the far-UV to the near-IR, we have to account for reddening and extinction towards the source. In our SED fits, we model the effects of reddening via the \citet{Fitzpatrick:1999PASP..111...63F} extinction curve with $R_V = 3.1$. The maximum line-of-sight extinction is $A_V \simeq 0.6$ \citep{schlegel,SF_extinction2011ApJ...737..103S}, corresponding to $E(B-V) \simeq 0.19$ for our adopted $R_V = 3.1$. 

With these assumptions, our SED fit has only four free parameters: (i) the temperature of the WD ($T_{WD}$; (ii) the mass of the donor ($M_2$); (iii) the reddening/extinction towards the source ($E(B-V)$); (iv) the distance towards the source ($d$). Given any combination of these parameters, we first calculate and sum the spectra of the two model component, apply the appropriate reddening and extinction, shift the model to the correct distance and then carry out synthetic photometry in the relevant filters. 

We estimate the optimal model parameters -- and their errors -- by carrying out a Markov-Chain Monte Carlo (MCMC) fit with {\tt dynesty}\footnote{https://zenodo.org/records/17268284} \citep{dynesty1} In doing so, we adopt the following simple and conservative uniform priors: 
\begin{enumerate}
    \item[(i)] $3.9 \leq \log{(T_{WD}~[\mathrm{K}])} \leq 4.7$
    \item[(ii)] $0.04 \leq M_2/M_{\odot} \leq 0.1$
    \item[(iii)] $0 \leq E(B-V) \leq 0.3$
    \item[(iv)] $1.0 \leq \log{(d~[\mathrm{pc}])} \leq 3.7$.
\end{enumerate}

The best fit achieved by this simple model is shown in Figure~\ref{fig:sed_fit} and the associated corner plot in Figure~\ref{fig:sed_fit_corner}. The fit is qualitatively and visually reasonable, but we have allowed for an intrinsic dispersion of $\sigma_{int} = 0.262$ to also make it statistically acceptable. This dispersion is added in quadrature to the observational error on each point ($\sigma_{obs,i}$), i.e. $\sigma^2_{total,i} = \sigma_{obs,i}^2+\sigma_{int}^2$. The value of $\sigma_{int}$ is set by requiring that a statistically acceptable fit should achieve $\chi^2_\nu \simeq 1$. Without the intrinsic dispersion term, our best-fitting model achieves $\chi_\nu^2 = 3.65$. Note that allowing for intrinsic dispersion term is conservative, in the sense that it increases the errors on the fit parameters. Of course, it is nevertheless important to ask whether the required level of intrinsic scatter is reasonable, and we will return to this point below. 

Overall, the best-fitting model does a reasonable job of reproducing the observed broad-band data. It suggests that a $15{,}000$~K WD dominates the SED from the far-UV band through the optical region. The donor only begins to contribute significantly in the $J$ band, before taking over in the $K_s$ band. The best-fit mass of the donor puts it firmly in the sub-stellar regime. 

The model parameters are all fairly well constrained by the data, although there are some degeneracies. All of these can be understood, however. For example, the WD temperature is highly correlated with $E(B-V)$, since an increase in $T_{WD}$ makes the model bluer, which can be counteracted by also  increasing the reddening. Similarly, the distance preferred by the model is correlated with both donor mass and WD temperature, because increasing the distance makes both the WD and secondary model components fainter. The only  way to counteract this WD [donor] flux decrease in the model is to increase the WD temperature [donor mass].

\section{Discussion}
\label{sec:discussion}

We have shown that a simple, physically motivated two-component model can provide a reasonable match to the far-UV through near-IR SED of ASKAP J1745-5051. The two components correspond to the accreting WD and its low-mass companion, with properties constrained by the mCV nature of this LPT. In particular, our model requires the donor mass, radius and temperature to fall on the CV donor sequence \citep[][KBP11]{Knigge2006}. The best-fit parameters -- $T_{WD} \simeq 14,500$~K, $M_2 \simeq 0.05~M_{\odot}$ (corresponding to $T_2 \simeq 1800$~K), $d \simeq 320$~pc and $E(B-V) \simeq 0.075$ -- are all quite reasonable compared to other period-bounce CVs. The X-ray (eROSITA/SRG) 
to optical (Gaia/G) flux ratio of ASKAP J1745-5051 ($\log{F_X/F_{opt}} \simeq 0.13$) is also consistent with that of polar-type mCVs \citep{schwope}, which can also be WD-dominated in the UV and optical bands. Thus the physical picture of the system developed here is fairly consistent internally. That said, several aspects of the model merit additional comments.

\subsection{Is the model good enough? The need for intrinsic dispersion}

As noted in Section~\ref{sec:SED_fitting}, we need to add an intrinsic dispersion of $\sigma_{int} \simeq 0.26$~mag (in quadrature) to the observational uncertainties in order for the model to achieve $\chi^2_\nu = 1$. Without any intrinsic dispersion, the best fitting model achieves $\chi^2_\nu = 3.65$. For a fit with $\nu = 5$ degrees of freedom (9 data points; 4 free parameters), this corresponds to a p-value of $p \simeq 0.003$. This p-value is the probability of finding a $\chi^2_\nu$ as large as this if both the model and the errors are correct. Formally, this amounts to a 3$\sigma$ rejection of the model. The {\em minimum} intrinsic dispersion needed -- which we define as that required to reach $p \simeq 0.05$ -- is $\sigma_{int, min} = 0.13$. Importantly, all of these fits -- from $\sigma_{int} = 0$~mag to $\sigma_{int} = 0.26$~mag -- produce almost identical best-fit parameters. What $\sigma_{int}$ does affect is the size of the inferred parameter uncertainties. As pointed out in Section~\ref{sec:SED_fitting}, allowing for intrinsic dispersion is conservative, in that it {\em increases} the parameter uncertainties.

In our view, the requirement for this level of intrinsic dispersion is not too concerning, either statistically or physically. Statistically, $\sigma_{int, min}$ is comparable to (or even smaller than)  most of the observational errors in Table~\ref{tab:obs_summary}. These errors are inevitably uncertain themselves, e.g. due to blending (which definitely affects the NIR and might affect the NUV). Physically, there are at least two plausible explanations for the fit residuals that would not fundamentally alter our conclusions. First, the source may exhibit variability that manifests as intrinsic dispersion in the fit. 
\footnote{Some polars exhibit distinct high and low states, which can differ in brightness by factors of a few in the UV \cite[e.g.][]{Gansicke1995}, but $\simeq 10-1000$ in X-rays \cite[e.g.][]{Ramsay2004, Schwope2007} $\simeq 5-50$ in the optical \cite[e.g.][]{wheatley98, Gerke2006} Our modelling implicitly assumes that all of the photometric data was obtained in at least roughly the same state. This seems likely, since the required intrinsic dispersion and residuals are much smaller than these numbers.}
In WD-dominated polars, a hot spot associated with the polar cap on the WD surface is one obvious way to generate variability on the orbital period \citep[e.g][]{Gansicke1995}. The amplitude of such variations can exceed a factor of two in the far-UV band, more than sufficient to account for the observed fit residuals (but not so large that the fit would be qualitatively affected). Another form of variability is short-time-scale stochastic ``flickering''. This is more commonly associated with disk-accreting systems \citep{2014MNRAS.438.1233S}, but is also seen in diskless polars \citep{2022MNRAS.516.5209I}. Second, since the WD is magnetic, its spectrum may exhibit significant cyclotron humps \citep{Cropper1990,Ferrario15}. These are not included in our model. The optical spectrum reported by \cite{Rose2026} does not reveal strong cyclotron features, but they could be more prominent in the ultraviolet \citep[e.g.][]{Gansicke2001} or nIR \citep[e.g.][]{Campbell2007} depending on the WD surface magnetic field strength. These features could thus produce localized residuals without changing the overall character of the fit. 

\subsection{Is the model too simple? The potential impact of missing model components} \label{sec:simpleModel}

A more significant concern is whether our model is too simple. In particular, it does not include any component associated with the accretion flow, assuming instead that the WD dominates everywhere shortward of the near-IR. In order to test the potential impact of this assumption, we have carried out fits with two additional models. These models are not intended to be physically realistic, but a simple way to assess whether/how a spectral component associated with the accretion flow might change our results. In the first type of model, we allow for an extra blackbody (BB) component described by temperature $T_{BB}$ and radius $R_{BB}$. In the second, we allow for an extra optically thick, truncated accretion disk. This disk is described as an area-weighted sum of blackbodies between inner radius $R_{inner}$ and outer radius $R_{disk} = 30~R_{WD}$. The outer disk radius corresponds to the expected tidal radius for a system with $P_{orb} \simeq 1.37$~hr, $M_{WD} = 0.8~M_{\odot}$ and $M_2 \simeq 0.05~M_{\odot}$. For simplicity, we fix the inclination at $i = 60^\circ$. The free parameters are then $R_{inner}$ and the accretion rate, $\dot{M}_{acc}$. Both of these are clearly just toy models for the accretion flow contribution. However, they provide a valuable check on the robustness of our simpler model.

Adding a single-temperature BB component to the model turns out to leave $T_{WD}$, $M_2$, $d$ and $E(B-V)$ almost unchanged. The optimal BB radius and temperature are $R_{BB} \simeq R_{WD}$ and $T_{BB} \simeq 7000$~K, making this component strongly sub-dominant everywhere. The fit quality also barely improves: for $\sigma_{int} = 0$, the raw $\chi^2$ value associated with our best-fit simple model is $\chi^2 = 18.25$, while that associated with the more complex model is $\chi^2 = 17.64$. Unsurprisingly, the BB parameters are essentially unconstrained. Even if we force the BB component to dominate in the FUV -- e.g. by setting $T_{BB} \simeq 50,000$~K and $R_{BB} \simeq 0.1~R_{WD}$, the remaining fit parameters change only modestly: $T_{WD} \simeq 11,000$~K, $M_2 \simeq 0.05~M_{\odot}$, $d \simeq 250$~pc and $E(B-V) \simeq 0.09$. Similarly, if we add a BB disk component to the model, the fit chooses to make this component negligible by driving $\dot{M}_{acc}$ towards zero. 

The only way to significantly alter the basic model parameters is to {\em force} the third component to dominate much of the SED, even at the cost of a worse fit. In such models, the distance is typically much larger, and the WD and donor parameters become poorly constrained. Most of these models are not physically plausible. For example, some require a very hot WD (to dominate in the far-UV despite a larger implied distance), while others simultaneously require a large distance and negligible reddening. However, with only 9 data points -- and 6 parameters for these more complex models -- we are clearly in the data-poor regime. In order to determine the WD and donor properties definitively, additional observations -- ideally far-UV and/or near-IR spectroscopy -- will be required. 

\subsection{The impact of systematic uncertainties affecting the near-IR brightness estimates}

As discussed in Section~\ref{sec:SED_construction}, our PSF-fitting near-IR brightness estimates are subject to considerable systematic errors. Depending on whether the PSF is modelled as a Gaussian or a Moffat profile, the inferred magnitudes can differ by up to 0.6 mag. Moreover, since the neighbouring star is significantly brighter than ASKAP J1745-5051 in the near-IR, even the significance of the detection itself may be questioned. However, treating the near-IR measurements as upper limits would simply tend to push the corresponding $M_2$ estimate even lower. In fact, this is exactly what happens when we repeat the fit after replacing the Gaussian-PSF near-IR brightness estimates with the fainter Moffat-PSF ones. This change moves the estimate of $M_2$ slightly downwards -- from $0.052~M_{\odot}$ to $0.049~M_{\odot}$ -- but otherwise leaves the fit largely unchanged. Our conclusions are therefore robust to these uncertainties.

\subsection{The WD temperature and implied accretion rate}

The WDs in CVs are thought to be accretion-heated \citep{Townsley_Bildstein2003}. Thus $T_{WD}$ can be used as a proxy for the accretion rate onto the WD, averaged over $10^3 - 10^5$~years and depending somewhat on the time since the most recent nova eruption of the system \citep{2009ApJ...693.1007T}. A WD temperature of $14{,}500$~K corresponds to $\dot{M}_{acc} \simeq 8 \times 10^{-11}~\mathrm{M_{\odot}~yr^{-1}}$. By comparison, the best-fit location on the CV donor sequence -- $M_2 \simeq 0.05~M_{\odot}$ -- corresponds to a mass-transfer rate of $\dot{M}_{acc} \simeq 3 \times 10^{-11}~\mathrm{M_{\odot}~yr^{-1}}$ and a WD temperature of $T_{WD} \simeq 11{,}200$~K. Allowing for both observational errors and theoretical uncertainties, we consider these values to be consistent.\footnote{Factors contributing to these ``theoretical uncertainties'' include the assumed WD mass, the poorly constrained CV angular momentum loss rates and the unknown time since the last nova eruption. They also include a possible far-UV contribution from the accretion-heated polar cap, which could reduce the $T_{WD}$ estimate to $\simeq 11{,}000$~K (cf. Section~\ref{sec:simpleModel}).}

Since ASKAP J1745-5051 most likely belongs to the polar class of mCVs, one theoretical uncertainty deserves special mention. KBP11's optimal donor and CV evolution sequence allows for angular momentum loss (AML) in excess of that expected from gravitational radiation alone. The need for this is well established in non-magnetic CVs. However, there is reasonable evidence that polar-type mCVs, in which the WD magnetic field is strong enough to prevent the formation of an accretion disk, may {\em not} require enhanced AML \citep{2009ApJ...693.1007T}. In these systems, WD temperatures tend to be systematically lower -- by perhaps a few thousand degrees -- than in non-magnetic CVs. This would increase the tension with the somewhat higher $T_{WD}$ favoured by the SED fit.

The closest confirmed polar-type mCV is AM~Her at $d \simeq 88$~pc \citep[e.g.][]{2020MNRAS.494.3799P}. The ratio of the space densities of ``normal'' polars and systems like ASKAP J1745-5051 is therefore roughly $(322/88)^3 \simeq 50$. However, this estimate neglects that the detection of LPTs requires the observer to lie within the narrow radio beam. Let us assume that the radio beam is a cone of half-opening angle $\rho$, centred on the axis of a dipolar magnetic field that is inclined by an angle $\alpha$ to the WD spin axis. The so-called ``beaming fraction'' -- the fraction of the sky swept out by the beam -- is then $f_{beam} = \sin{\alpha}\sin{\rho}$ for $\alpha \geq \rho$, and $f_{beam} = [1 - \cos{(\alpha + \rho)}]/2$ for $\alpha < \rho$ \citep[e.g.][]{Lorimer_Kramer:2004hpa..book.....L}. Based on Supplementary Table 1 in \citet{Rose2026}, we estimate that the typical orbital phase width of the radio pulses in ASKAP J1745-5051 is $\Delta\phi \simeq 0.07$. For this pulse duty cycle, $\rho$ depends on the unknown magnetic inclination and on how closely our line of sight passes to the magnetic axis \citep{Gil:1984A&A...132..312G}. If our line of sight crosses the centre of the beam, the observed pulse width implies $\rho \simeq 13^\circ$ for an orthogonal rotator ($\alpha = 90^\circ$), but only $\rho \simeq 3^\circ$ for $\alpha = 15^\circ$. The corresponding beaming fractions are $f_{beam} \simeq 0.2$ and $f_{beam} \simeq 0.015$, respectively. A line of sight that only grazes the edge of the beam allows wider beams, up to $\rho \simeq 25^\circ$ and $f_{beam} \simeq 0.4$ for $\alpha = 90^\circ$. Plausible geometries therefore span $f_{beam} \approx 0.02 - 0.4$, with the smallest values corresponding to small magnetic inclinations. If the magnetic inclination is modest, the {\em intrinsic} space density of systems like ASKAP J1745-5051 could thus be more than an order of magnitude higher than implied by its distance. Thus the percentage of mCVs that {\em produce} pulsed radio emission could actually be quite high. In this scenario, the reason we do not observe most of them as LPTs is simply that we are not aligned with the radio beam.

\section{Summary}
\label{sec:summary}

We have constructed and modelled the far-UV through near-IR SED of the recently discovered LPT and mCV ASKAP J1745-5051. We find that the SED can be reasonably described by a simple two component model: an accretion-heated WD and a low-mass donor that is constrained to lie on the standard CV evolution track. Based on this analysis, our main results are as follows.
\begin{enumerate}
    \item[(i)] The far-UV through optical flux can be explained by emission from the $\simeq 15{,}000$~K WD. 
    \item[(ii)] A sub-stellar donor of mass $M_2 \simeq 0.05~M_{\odot}$ begins to contribute in the near-IR and dominates the $K_s$ band. With such a small donor mass, the system is likely to already have evolved past the period minimum and is likely to be a ``period bouncer''.
    \item[(iii)] Our SED fit suggests a distance towards ASKAP J1745-5051 of $d \simeq 320 \pm 70$~pc. 
    \item[(iv)] The space density implied by this distance corresponds to $\simeq 2$\% of the mCV space density, but the beaming fraction may well be small. If so, the true percentage of LPTs among mCV could be far larger.
\end{enumerate}

\section*{Acknowledgments}

We are grateful to Kovi Rose, both for their helpful responses to our questions and for making most of their data and analysis scripts publicly available via github and zenodo. The code associated with this paper was created with Generative AI (mostly Anthropic's Claude), but always with a human -- in this case the lead author -- in the loop. The analysis and interpretation also benefitted from discussions with Claude. Responsibility for the results -- and for any errors -- rests solely with the authors, of course. SS is supported by STFC grants ST/T000244/1 and ST/X001075/1. DDM is supported by INAF,  AF2022 FANS and AF2024 PULSE-X projects. NCS acknowledges support from the Science and Technology Facilities Council (STFC) grant ST/X001121/1.

This work has made use of data from the European Space Agency (ESA) mission
{\it Gaia} (\url{https://www.cosmos.esa.int/gaia}), processed by the {\it Gaia}
Data Processing and Analysis Consortium (DPAC,
\url{https://www.cosmos.esa.int/web/gaia/dpac/consortium}). Funding for the DPAC
has been provided by national institutions, in particular the institutions
participating in the {\it Gaia} Multilateral Agreement. This research has made
use of the SVO Filter Profile Service
(\url{http://svo2.cab.inta-csic.es/theory/fps/};
\citealt{2012ivoa.rept.1015R, 2020sea..confE.182R, 2024A&A...689A..93R}).
This work made use of the following software: \texttt{NumPy}
\citep{2020Natur.585..357H}, \texttt{SciPy} \citep{2020NatMe..17..261V},
\texttt{Matplotlib} \citep{2007CSE.....9...90H}, \texttt{Astropy}
\citep{2013A&A...558A..33A, 2018AJ....156..123A, 2022ApJ...935..167A},
\texttt{astroquery} \citep{2019AJ....157...98G}, \texttt{photutils}
\citep{photutils_v3.0.0}, \texttt{synphot} \citep{2018ascl.soft11001S},
\texttt{stsynphot} \citep{2020ascl.soft10003S}, \texttt{dust\_extinction}
\citep{2024JOSS....9.7023G}, \texttt{dynesty} (\citealt{dynesty1};
\citealt{dynesty_v3.0.0}) and \texttt{corner} \citep{2016JOSS....1...24F}.

\bibliographystyle{aasjournal}
%\bibliographystyle{apsrev4-1}

% You should give the same name for your .bbl as your main .tex
% since it is a requirement for posting on ArXiv.
\newpage
\bibliography{oja_template}

% \begin{appendix}

% \section{Appendix 1}
% \label{ap:ap}

% \end{appendix}

\end{document}